\documentclass[pra,twocolumn,showpacs,groupedaddress,superscriptaddress,aps,10pt]{revtex4-1}
\usepackage{bm,graphicx,amsmath}
\usepackage{placeins}
\usepackage{amssymb}
\usepackage{amsmath}
\usepackage{epsfig}
\usepackage{amssymb}
\usepackage{color}
\usepackage[colorlinks,linkcolor=blue,citecolor=blue]{hyperref}
\usepackage{subfigure}

\newcommand{\etal}{{\it{et al.~}}}

\begin{document}

\title{Conical refraction healing after partially blocking the input beam}
\date{\today}

\author{Alex Turpin} \email{Corresponding author: alejandro.turpin@uab.cat}
\affiliation{Departament de F\'isica, Universitat Aut\`onoma de Barcelona, Bellaterra, E-08193, Spain}
\author{Yury V. Loiko}
\affiliation{Aston Institute of Photonic Technologies, School of Engineering and Applied Science Aston University, Birmingham, B4 7ET, UK}
\author{Todor K. Kalkandjiev}
\affiliation{Departament de F\'isica, Universitat Aut\`onoma de Barcelona, Bellaterra, E-08193, Spain}
\affiliation{Conerefringent Optics SL, Avda. Cubelles 28, Vilanova i la Geltr\'u, E-08800, Spain}
\author{Ram\'on Corbal\'an}
\affiliation{Departament de F\'isica, Universitat Aut\`onoma de Barcelona, Bellaterra, E-08193, Spain}
\author{Jordi Mompart}
\affiliation{Departament de F\'isica, Universitat Aut\`onoma de Barcelona, Bellaterra, E-08193, Spain}

\begin{abstract}
In conical refraction, when a focused Gaussian beam passes along one of the optic axes of a biaxial crystal it is transformed into a pair of concentric bright rings at the focal plane. We demonstrate both theoretically and experimentally that this transformation is hardly affected by partially blocking the Gaussian input beam with an obstacle. We analyze the influence of the size of the obstruction both on the transverse intensity pattern of the beam and on its state of polarization, which is shown to be very robust. 
\end{abstract}

\date{\today }
%
%
\maketitle
\section{Introduction}
\label{intro}

Gaussian beams are the most well known solution of the paraxial wave equation. They are form-invariant beams, i.e., the form of their transverse intensity pattern does not change upon propagation, apart from a scaling factor. Durnin \etal \cite{durnin:1987:prl} reported another solution of the paraxial wave equation, the Bessel beams, which are completely invariant upon propagation. In other words, both the transverse intensity profile and scale of Bessel beams remain unchanged as it propagates, i.e., Bessel beams are diffraction-free beams. One of the main features of Bessel beams is that they self-reconstruct after an obstacle, being this effect known as self-healing. 

Recently, there has been a great interest in the study of the self-healing effect appearing in Bessel beams \cite{sato:2011:josaa,cai:2014:pra} and other diffraction-free beams including Airy beams \cite{demetrios:2007:prl,demetrios:2008:oe} and Pearcey beams \cite{dennis:2012:oe}, or other exotic beams such as helico-conical beams \cite{jpt:2013:ol} as well as Mathieu and Webber beams \cite{zhang:prl:2012}. The major advantage of self-healing beams is that they can be used through turbulent media \cite{demetrios:2008:oe} and that they are ideal candidates for particle manipulation at different planes \cite{dholakia:2002:nature,dholakia:2008:nat_photon} and in microscopy \cite{rohrbach:2010:nat_photon}. 

\begin{figure}[htb]
\centering
\includegraphics[width=1 \columnwidth]{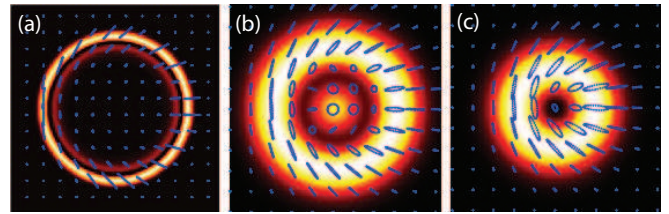}
\caption{(Color online) Transverse intensity patterns and state of polarization at the focal plane of CR beams numerically calculated from Eqs.~(\ref{E_input})-(\ref{B1_CR}) for a left handed circularly polarized Gaussian input beam for (a) $\rho_0 \equiv R_0/w_0 = 10$, (b) $\rho_0 = 1.5$ and (c) $\rho_0 = 0.92$. $R_0$ is the geometrical CR ring radius and $w_0$ the waist radius of the focused Gaussian input beam.}
\label{fig1}
\end{figure}

In conical refraction (CR) \cite{belsky:1978:os,berry:2007:po,todor:2008:spie}, when a circularly polarized focused Gaussian beam propagates through a biaxial crystal parallel to one of the optic axes it appears transformed into a pair of concentric bright rings split by a null intensity (Poggendorff) ring at the focal plane, see Fig.~\ref{fig1}(a). At this plane, the state of polarization of the rings is linear at every point, with the azimuth rotating continuously so that every two diametrically opposite points have orthogonal polarizations. This non-factorization of the state of polarization makes CR beams a very particular type of vector beams \cite{turpin_stokes:2015:oe}. Along the axial direction, the transverse intensity pattern of the CR beam becomes more involved and far enough from the focal plane an axial (Raman) spot is formed. The above named features of CR beams are valid under the assumption that the CR geometrical ring radius $R_0$ is much bigger than the waist radius of the focused input beam $w_0$, i.e., assuming that $\rho_0 \equiv R_0/w_0 \gg 1$. For smaller values of $\rho_0$, different light structures are formed \cite{peet:2010:oe,loiko:2013:ol,turpin:2014:ol}. 

CR has been reported as an efficient tool to generate Bessel beams \cite{kazak:1999:qe,king:2001:oc,phelan:2009:oe}. The relation of CR with Bessel beams suggests that even if the input Gaussian beam is blocked partially, the CR beam may only be slightly affected. 
In this work we address the question of the CR healing of a Gaussian input beam in the presence of an obstruction. We investigate the reconstruction of the transverse intensity pattern of the CR beams at different propagation distances and also of their state of polarization. We show that, even for relatively large obstacles, the CR beams keep their annular shape, state of polarization and dark optical singularities. The work is organized as follows: the theory of CR is presented in Section~\ref{theory}. In Section~\ref{rho0big} we describe our experimental set-up and report healing of conically refracted Gaussian beams after an obstruction for $\rho_0 \gg 1$. In Section~\ref{rho0small} we compare the reconstructing behavior of the CR beam for $\rho_0 = 0.92$ and $\rho_0 \gg 1$ and discuss the differences between the two
cases. Finally, our main results are summarized in Section~\ref{conclusions}.

\section{Paraxial theory of conical refraction}
\label{theory}

In what follows, we normalize our coordinates components to the beam waist $w_0$ and Rayleigh range $z_R$, i.e., $x \rightarrow x/w_0$, $y \rightarrow y/w_0$ and $z \rightarrow z/z_R$. 
Additionally, note that when a light beam propagates through a biaxial crystal (with three different refractive indices $n_x$, $n_y$ and $n_z$) along one of the optic axis, there is a degeneracy on the refractive indices, i.e., $n_x = n_z = n_y \equiv n$ \cite{berry:2007:po}. Therefore, in the forthcoming equations only $n$ needs to be taken into account (for the crystal used in the experiments described below $n \approx 2$ \cite{todor:2008:spie}).

Let's consider an input beam with transverse electric field 
\begin{equation}
\mathbf{E}(\mathbf{r}) = E_x \mathbf{e}_x + E_y \mathbf{e}_y,
\label{E_input}
\end{equation}
where $\mathbf{e}_x$ and $\mathbf{e}_y$ are the unit vectors in Cartesian coordinates, and corresponding Fourier transform 
\begin{eqnarray}
\mathbf{A}(\mathbf{k}) &=& A_x(\mathbf{k}) \mathbf{e_x} + A_y(\mathbf{k}) \mathbf{e_y}, \\ \label{FT_vector}
A_{x}(\mathbf{k}) &=& \frac{1}{(2 \pi)^2} \iint\limits_{-\infty}^{\infty}
E_{x}(\mathbf{r})  e^{-i \mathbf{k} \cdot \mathbf{r}} dx dy,\\ \label{FT_x}
A_{y}(\mathbf{k}) &=& \frac{1}{(2 \pi)^2} \iint\limits_{-\infty}^{\infty}
E_{y}(\mathbf{r})  e^{-i \mathbf{k} \cdot \mathbf{r}} dx dy. \label{FT_y}
\end{eqnarray}

It can be shown that for a low birefringent biaxial crystal the beam evolution behind the biaxial crystal can be described as follows \cite{turpin_general:2015_arxiv}: 
\begin{eqnarray}
E_{x} &=& B_{1, x}(\mathbf{r},\rho_0) + B_{0, y}(\mathbf{r},\rho_0),
\label{E_CRx}\\
E_{y} &=& B_{0, x}(\mathbf{r},\rho_0) + B_{1, y}(\mathbf{r},-\rho_0). 
\label{E_CRy}
\end{eqnarray}
with two main integrals 
\begin{widetext}
\begin{eqnarray}
B_{0, \alpha}(\mathbf{r},\rho_0)&=& \frac{i}{(2 \pi)^2} \iint\limits_{-\infty}^{\infty} e^{-i (\mathbf{k} \cdot \mathbf{r} - \frac{z}{2n} k^2))} \frac{k_y}{k} \sin \left( \rho_0 k \right) A_{\alpha}(\mathbf{k}) d k_x d k_y,\label{B0_CR} \\ 
B_{1, \alpha}(\mathbf{r},\rho_0) &=&
\frac{1}{(2 \pi)^2} \iint\limits_{-\infty}^{\infty} e^{-i (\mathbf{k} \cdot \mathbf{r}  - \frac{z}{2n} k^2)} \left( \cos \left( \rho_0 k \right) + i \frac{k_x}{k} \sin \left( \rho_0 k \right) \right) A_{\alpha}(\mathbf{k}) d k_x d k_y,~\label{B1_CR}
\end{eqnarray}
\end{widetext}
where $\alpha=x,y$.

In this work, we consider a Gaussian input beam with electric field $\mathbf{E_{\rm{in}}} \propto \exp \left( -(x^2+y^2)\right)\mathbf{e}_{\rm{in}}$, where $\mathbf{e}_{\rm{in}}$ is the state of polarization of the beam. Fig.~\ref{fig1} shows the transverse intensity patterns and state of polarization at the focal plane of CR beams numerically calculated from Eqs.~(\ref{E_input})-(\ref{E_CRy}) for a left handed circularly polarized Gaussian input ($\mathbf{e}_0 = \frac{1}{\sqrt{2}} (1,i)$) beam for (a) $\rho_0 = 10$, (b) $\rho_0 = 1.5$ and (c) $\rho_0 = 0.92$. As it can be observed, both the transverse intensity pattern and the polarization of the beams depend on $\rho_0$, as it has been shown in detail in \cite{turpin_stokes:2015:oe}. For $\rho_0 = 10 \gg 1$, all the features of the CR beam described above can be observed, namely: the transverse intensity pattern is formed by two bright rings split by the Poggendorff dark ring and every two opposite points are orthogonally linearly polarized. Note that under the condition $\rho_0 \gg 1$, the properties of the CR beam are unaffected by slightly different values of $\rho_0$. 
For $\rho_0 = 1.5$ the inner ring collapses into a point and the region separating the central spot and the outer ring possesses low but non-null intensity. For $\rho_0 = 0.92$ the pattern is doughnut-like, being the central dark spot an exact null-intensity point, as reported in \cite{loiko:2013:ol}. In both cases the state of polarization is non-homogeneously distributed, including regions with elliptical states of different azimuth and ellipticity. Nonetheless, at the edge of the ring the polarization distribution is linear with azimuth rotating an angle $\phi$ for a full turn along the ring. The dependence of the CR transverse intensity pattern with $\rho_0$ has been analyzed in detail in \cite{peet:2010:oe}. 

\section{Healing of conically refracted Gaussian beams for $\rho_0 \gg 1$}
\label{rho0big}

\begin{figure}[htb]
\centering
\includegraphics[width=1 \columnwidth]{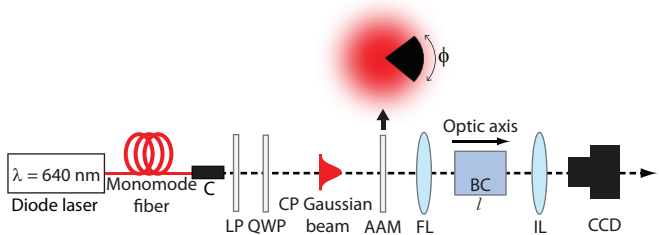}
\caption{(Color online) Experimental set-up. A Gaussian beam is obtained from a diode laser coupled to a monomode fiber. A collimator (C) is used to change the beam waist radius $w_0$ of the Gaussian beam. A linear polarizer (LP) and a quarter waveplate (QWP) are used to fix the state of polarization of the Gaussian beam to circular. Azimuthal angular amplitude masks (AAM) characterized by their closing angle $\phi$ are used to block a sector of the Gaussian beam. The obtained beam passes through the biaxial crystal (BC) along one of its optic axes and a CCD camera and an imaging lens (IL) record the transverse intensity pattern at different planes.}
\label{fig2}
\end{figure}

In this Section we analyze the reconstruction of conically refracted Gaussian beams under the condition $R_0 \gg w_0$ when an obstruction of closing angle $\phi$ blocks an azimuthal part of the input beam. We consider first the approximation $R_0 \gg w_0$ since it is the commonly used configuration in most experimental arrangements. 
Fig.~\ref{fig2} shows our experimental set-up. We obtain a circularly polarized Gaussian beam at $640\,\rm{nm}$ from a diode laser coupled to a monomode fiber by utilizing, a linear polarizer (LP) and a quarter wave-plate (QWP). As an obstruction, we use amplitude angular masks (AAM). They block an azimuthal sector of angle $\phi$ of the Gaussian beam at the exit of the collimator. The AAM are made by printing the desired 2D pattern over a transparent sheet of plastic. The beam, whose waist radius $w_0$ can be adjusted by means of the collimator, passes through a biaxial crystal and parallel to one of the optic axes and a CCD camera combined with an imaging lens (IL) records the transverse intensity pattern of the CR beam at different planes. As a biaxial crystal we use a commercially available (CROptics) $\rm{KGd(WO_4)_2}$ of length $l=28\,\rm{mm}$, conicity $\alpha = 16.9$~mrad and, therefore, $R_0 = l \alpha \approx 475\,\rm{\mu m}$. The polished entrance surfaces of the crystal (cross-section $6 \times 4~\rm{mm^2}$) have parallelism with less than 10~arc sec of misalignment, and they are perpendicular to one of the two optic crystal axes within $1.5\,\rm{mrad}$ misalignment angle for the central wavelength $633\,\rm{nm}$. 
The crystal is mounted over a micro-positioner. Fine alignment is performed by modifying independently the $\theta$ and $\phi$ angles of the micro-positioner and observing intensity pattern at the focal plane with the CCD camera. A cylindrically symmetric pattern is obtained when the input beam of circular polarization passes exactly along one of the optic axis of the BC. By reducing the focused beam waist down to $w_0 = 44\,\rm{\mu m}$, we have obtained a $\rho_0$ parameter up to $10.75$. 

\begin{figure}[htb]
\centering
\includegraphics[width=1 \columnwidth]{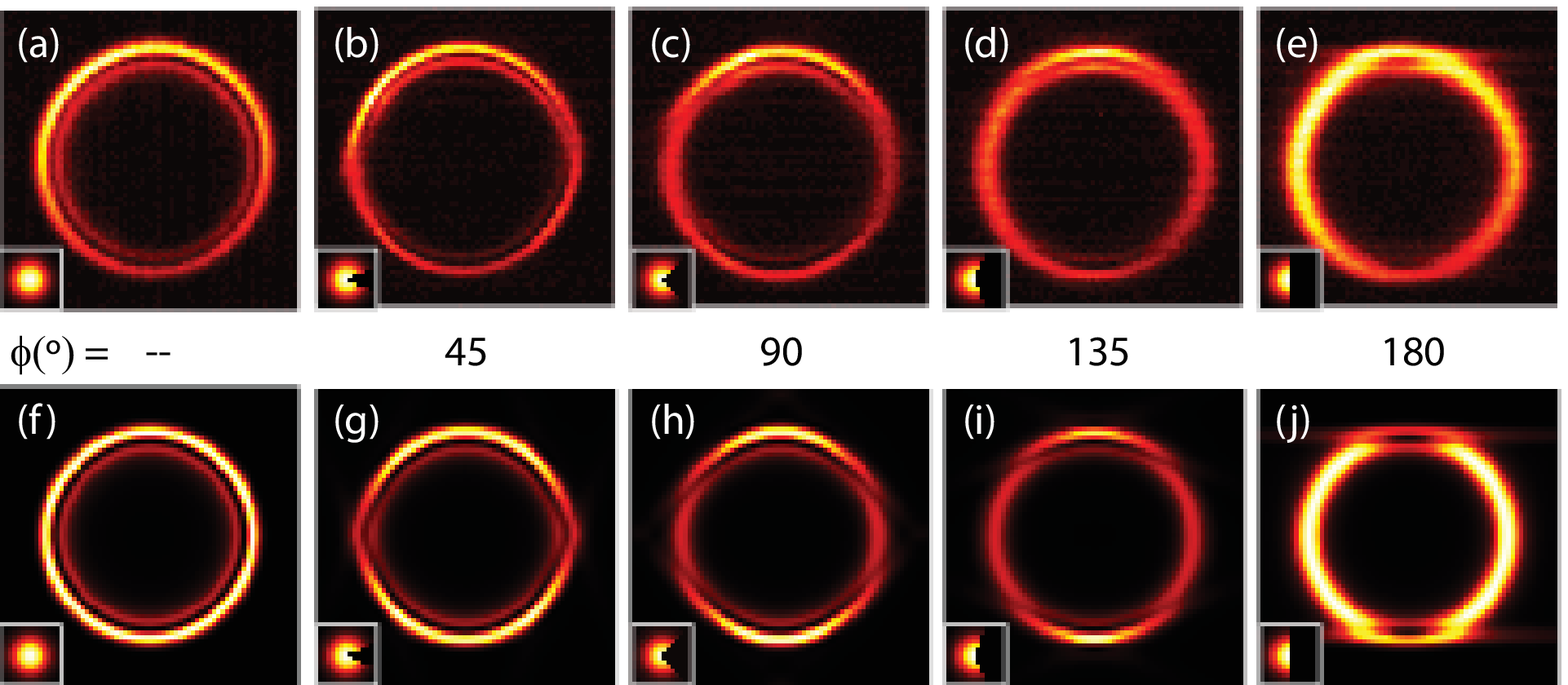}
\caption{(Color online) Transverse intensity patterns at the focal plane ($z=0$) obtained with an obstruction of closing angle (a,f) $\phi = 0^{\circ}$, (b,g) $\phi = 45^{\circ}$, (c,h) $\phi = 90^{\circ}$, (d,i) $\phi = 135^{\circ}$ and (e,j) $\phi = 180^{\circ}$ placed before the biaxial crystal. (a,f) Intensity pattern obtained in the absence of obstruction. First row: experimental results ($\rho_0^{\rm{exp}} = 10.75$). Second row: numerical calculations obtained from Eqs.~(\ref{E_input})--(\ref{E_CRy}) ($\rho_0^{\rm{th}} = 10$).}
\label{fig3}
\end{figure}

Fig.~\ref{fig3} shows both the experimental (top row) and numerically calculated (bottom row) transverse intensity pattern at the focal plane for obstructions of (a,f) $\phi = 0^{\circ}$, (b,g) $45^{\circ}$, (c,h) $90^{\circ}$, (d,i) $135^{\circ}$ and (e,j) $180^{\circ}$. Insets represent the obstructed input beam just behind the AMM. When a relatively small obstruction angle is considered ($\phi=45^{\circ}$), the transverse intensity pattern is almost unaffected as compared with the case without obstruction. In this case, the Poggendorff dark ring and the two ring-like structures are clearly visible. The rings are mirror-symmetric with respect to the horizontal axis but they have a maximum at their top and bottom regions. 
As $\phi$ increases the CR rings become asymmetric and the outer ring breaks into two boomerang-like lobes such that no complete Poggendorff dark ring is appreciable. For $\phi=180^{\circ}$, i.e., when the AAM blocks half of the input beam, the CR transverse intensity pattern is formed by a wide single ring with two dark singularities in the upper and top regions of the ring. 

\begin{figure}[htb]
\centering
\includegraphics[width=1 \columnwidth]{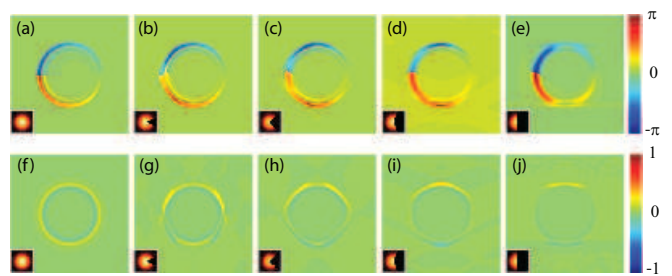}
\caption{(Color online) Numerically calculated 2D density plots of the azimuth $\epsilon$ (first row) and the ellipticity $\beta$ (second row) of conically refracted beams at the focal plane for $\rho_0 = 10$ when the input Gaussian is blocked by azimuthal obstructions of (b,g) $\phi = 45^{\circ}$, (c,h) $\phi = 90^{\circ}$, (d,i) $\phi = 135^{\circ}$ and (e,j) $\phi = 180^{\circ}$ is placed before the biaxial crystal. (a,f)  2D density plots of $\epsilon$ and $\beta$ in the absence of the obstruction.}
\label{fig5}
\end{figure}

The healing mechanism provided by the biaxial crystal to reconstruct the CR beam can be understood in terms of the wave-vector splitting within the crystal. Every plane-wave is described by a certain wave-vector $\mathbf{k} = \mathbf{k}_{\parallel} + \mathbf{k}_{\perp}$, with $\mathbf{k}_{\perp} = |k_{\perp}| (\cos\phi_k, \sin\phi_k)$. The biaxial crystal splits every plane wave into two new plane-waves. At the focal plane, these two plane-waves are refracted at positions on the ring characterized by their azimuthal angle $\varphi = \phi_{k}$ and $\varphi = \phi_{k}+\pi$ \cite{turpin_ebs:2013:oe}. As a consequence, when one azimuthal sector of the Gaussian beam is blocked, the azimuthally opposite sector partially compensates the absence of the blocked sector. For this reason, even when half of the input beam is blocked, a ring-like structure can be formed after passing through the biaxial crystal. 

This mechanism explains why a single bright ring without Poggendorff splitting is obtained when half of the input beam is blocked with the AAM, i.e., when $\phi = 180^{\circ}$. The two bright rings with Poggendorff splitting appear as an interference of plane waves going to a particular azimuthal point of the CR ring from opposite sectors of the input beam. In contrast, if there are no other waves coming to the corresponding opposite points of the CR ring pattern at the focal plane, there is no interference and a only a single ring is observed. 
In Figs.~\ref{fig3}(b)--(e) it is clearly visible an increase of the azimuthal sector of single bright ring and shrink of the double bright rings' domain with the AMM closing angle $\phi$. Note that the azimuthal sector occupied by the double bright rings is larger than the angular sector of the AMM, $\phi$, because of the diffraction of the input beam at the edges of AMM dark sector.

Now we turn to analyze the healing of the state of polarization of the reconstructed CR beams. The standard tool to analyze the state of polarization of a light beam is the Stokes vector: $S = (S_0,S_1,S_2,S_3)$. For an electric field $\mathbf{E} = (E_x, E_y)$ with intensity $I$ the Stokes parameters read \cite{Goldstein}:
\begin{eqnarray}
S_0 &=& I = \left| E_x \right|^2 + \left| E_y \right| ^2,\\
\label{S0}
S_1 &=& I_{0^{\circ}} - I_{90^{\circ}} = \left| E_x \right|^2 - \left| E_y \right|^2,\\
\label{S1}
S_2 &=& I_{45^{\circ}} - I_{135^{\circ}} = 2 \rm{Re} \left[ E_x^* E_y \right],\\
\label{S2}
S_3 &=& I_{R} - I_{L} = 2 \rm{Im} \left[ E_x^* E_y \right],
\label{S3}
\end{eqnarray}
where $I_{\Phi}$ ($\Phi=0^{\circ},45^{\circ},90^{\circ},135^{\circ}$) indicates the intensity of linearly polarized light with azimuth $\Phi$, and $I_R$ and $I_L$ indicate the intensity of right- and left-handed circularly polarized light, respectively. In what follows, we use equations normalized to $E^2$, i.e., we consider $I = E^2 = 1$. These definitions show that $S_0$ account for the intensity of the light beam, $S_1$ measures the amount of light which is linearly polarized (LP) in the vertical/horizontal basis, $S_2$ does the same but with the diagonal basis and $S_3$ relates the state of polarization in the right- and left- circularly polarized (CP) basis. The following equations show how the Stokes parameters account for the azimuth $\epsilon$ and ellipticity $\beta$ of the polarization ellipse \cite{Goldstein}:
\begin{eqnarray}
\epsilon &=& \frac{1}{2} \arctan{\left( \frac{S_{2}}{S_{1}} \right)},
\label{phi_stokes} \\
\beta &=& \frac{1}{2} \arctan{\left( \frac{S_{3}}{\sqrt{S_{1}^2+S_{2}^2}} \right)}.
\label{beta_stokes}
\end{eqnarray}
The values of $\epsilon$ and $\beta$ at the focal plane numerically calculated for obstructions of (a,f) $\phi = 0^{\circ}$, (b,g) $45^{\circ}$, (c,h) $90^{\circ}$, (d,i) $135^{\circ}$ and (e,j) $180^{\circ}$ are presented in Fig.~\ref{fig5}. Insets represent the obstructed input beam. Ideally, as $\rho_0 \rightarrow \infty$, $\beta \rightarrow 0$. Since in our numerical simulations we consider $\rho_0 = 10$, the state of polarization of the CR rings is slightly elliptical rather than purely linear. As it can be appreciated, in general, the polarization structure of the CR beam is maintained for all the values of $\phi$, i.e., every two diametrically opposite points of the light structure are orthogonally polarized. 

\begin{figure*}[htb]
\centering
\includegraphics[width=2 \columnwidth]{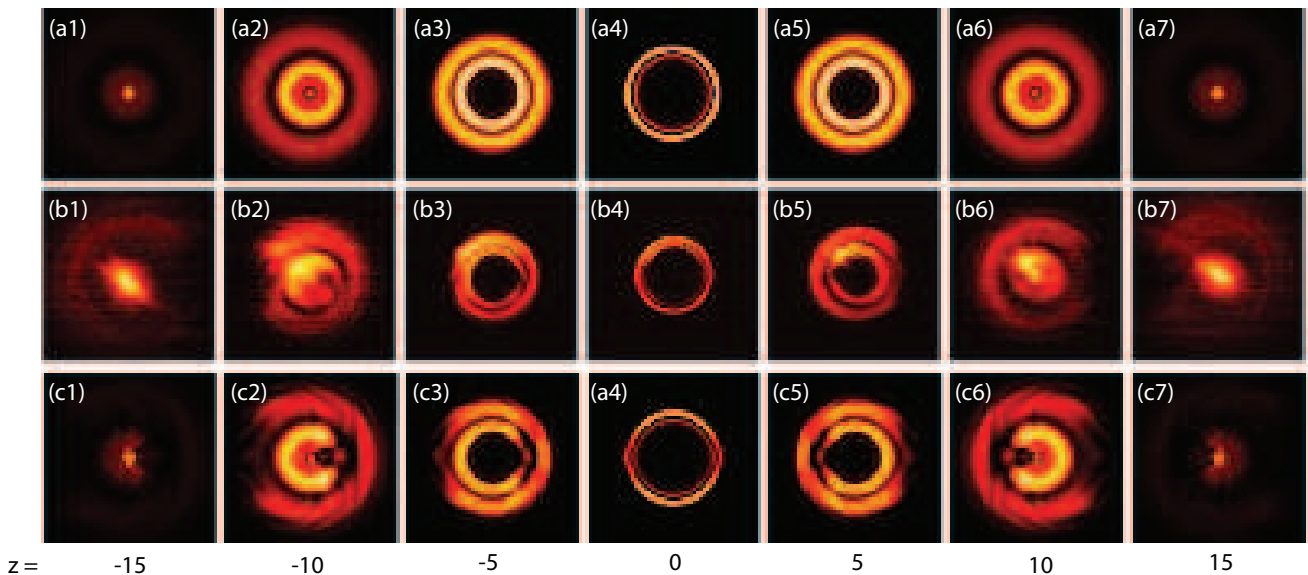}
\caption{(Color online) Experimental (second row) and numerically calculated with Eqs.~(\ref{E_input})--(\ref{E_CRy}) (third row) transverse intensity patterns along the axial direction $z$ for an obstruction with $\phi=45$ for $\rho_0 = 10$. The first row shows the numerically calculated transverse intensity patterns obtained in the absence of the obstruction. $z$ is measured in units of the Rayleigh range, which for the Gaussian beam used in our experiments ($w_0 = 44\,\rm{\mu m}$) is $z_R = 9.5\,\rm{mm}$.}
\label{fig4}
\end{figure*}

In Fig.~\ref{fig4} we show the evolution of the transverse intensity pattern along the axial direction for a blocking sector of $\phi=45^{\circ}$. Top row shows the transverse intensity patterns in the absence of the blocking mask, while middle and bottom rows are the experimental and numerically calculated transverse intensity patterns for the obstructed Gaussian beam. 
Near the focal plane ($z=0$) the transverse intensity pattern resembles the pattern obtained without obstruction. In contrast, far enough of the focal plane it can be appreciated a perturbation of the CR transverse intensity pattern that resembles the considered obstruction. 

\section{Reconstruction properties of conically refracted Gaussian beams for $\rho_0 \approx 1$}
\label{rho0small}

\begin{figure}[htb]
\centering
\includegraphics[width=1 \columnwidth]{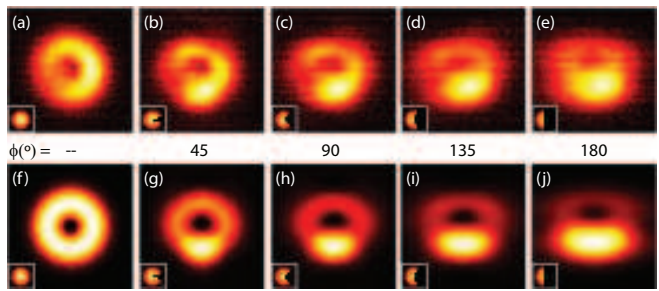}
\caption{(Color online) Transverse intensity patterns at the focal plane ($z=0$) obtained when an obstruction of angle (b,g) $\phi = 45^{\circ}$, (c,h) $\phi = 90^{\circ}$, (d,i) $\phi = 135^{\circ}$ and (e,j) $\phi = 180^{\circ}$ is placed before the biaxial crystal. (a,f) Transverse intensity pattern in the absence of obstruction. First row: experimental results ($\rho_0^{\rm{exp}} = 1.04$). Second row: numerical calculations obtained from Eqs.~(\ref{E_input})--(\ref{E_CRy}) ($\rho_0^{\rm{th}} = 0.92$).}
\label{fig6}
\end{figure}

As discussed in Section~\ref{theory}, the CR beam depends strongly on the value of the control parameter $\rho_0$. For $\rho_0 = 0.92$ the transverse intensity pattern at the focal plane forms a doughnut-like light structure with a null intensity point at the beam center \cite{loiko:2013:ol}, see Fig.~\ref{fig6}(a). Along the axial direction, the intensity at the beam center is no longer zero and the beam forms an optical bottle. This value of $\rho_0$ is particularly interesting because the polarization distribution of the light ring has points with different $\beta$ and $\epsilon$ and the beam forms a Poincare beam, i.e., a beam possessing points with all the polarization states of the Poincare sphere \cite{alonso:2010:oe}. In what follows we discuss the reconstruction of a conically refracted Gaussian beam for $\rho_0 = 0.92$ after an obstruction, analogously to what has been performed above for $\rho_0 \gg 1$. For the incoming experiments we have used the same set-up as in Fig.~\ref{fig2} but with a $2.3\,\rm{mm}$ long $\rm{KGd(WO_4)_2}$ (therefore $R_0 = 39\,\rm{\mu m}$) crystal and a waist radius of $w_0 = 44\,\rm{\mu m}$.

Fig.~\ref{fig6} shows both the experimental (top row) and numerically calculated (bottom row) transverse intensity pattern at the focal plane for obstructions of (a,f) $\phi = 0^{\circ}$, (b,g) $45^{\circ}$, (c,h) $90^{\circ}$, (d,i) $135^{\circ}$ and (e,j) $180^{\circ}$. Insets represent the obstructed input beam for $\rho_0 = 0.92$. In contrast to the case of $\rho_0 \gg 1$, when a relatively small obstruction angle is considered ($\phi=45^{\circ}$), the transverse intensity pattern is substantively different with respect to the case with no obstruction. In this case, a maximum of intensity appears at the bottom part of the ring and, therefore, the intensity pattern is mirror symmetric with respect to the vertical axis. As $\phi$ increases, the intensity in the bottom part of the light structure becomes stronger than in the top part. For all the values of $\phi$, an intensity minimum can be observed but its position moves in the vertical direction as $\phi$ increases. However, note that even when half of the beam is blocked by the obstruction, the dark singularity is preserved. A similar behavior has been reported for a linearly polarized Bessel beam \cite{cai:2014:pra}. 

\begin{figure}[htb]
\centering
\includegraphics[width=1 \columnwidth]{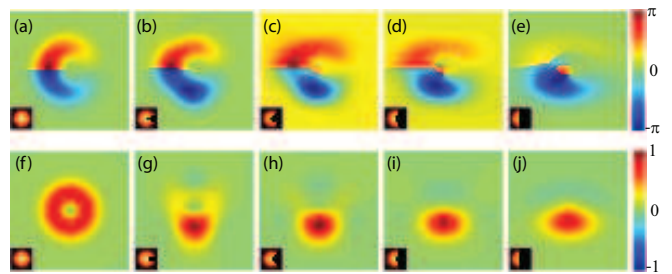}
\caption{(Color online) 2D density plots of the azimuth $\epsilon$ (first row) and the ellipticity $\beta$ (second row) of conically refracted beams at the focal plane for $\rho_0 = 0.92$ when the input Gaussian is blocked by azimuthal obstructions of (b,g) $\phi = 45^{\circ}$, (c,h) $\phi = 90^{\circ}$, (d,i) $\phi = 135^{\circ}$ and (e,j) $\phi = 180^{\circ}$ is placed before the biaxial crystal. (a,f)  2D density plots of $\epsilon$ $\beta$ in the absence of obstruction.}
\label{fig8}
\end{figure}

With respect to the state of polarization of the CR beams at the focal plane, see Fig.~\ref{fig8}, we have observed that there is a tendency to preserve the polarization structure of the CR beam without obstruction: the CR beam is elliptically polarized and the ellipticity changes radially similarly to the intensity pattern. At the edges of the beam, $\beta \rightarrow 0$ and the characteristic CR polarization distribution is recovered. However, since the transverse intensity pattern is very affected by the presence of an obstruction, these features of the state of polarization are lost for large enough values of $\phi$. In particular, $\beta$ losses its doughnut-like shape, while $\epsilon$ is kept quite stable up to $\phi = 135^{\circ}$. 

\begin{figure*}[htb]
\centering
\includegraphics[width=2 \columnwidth]{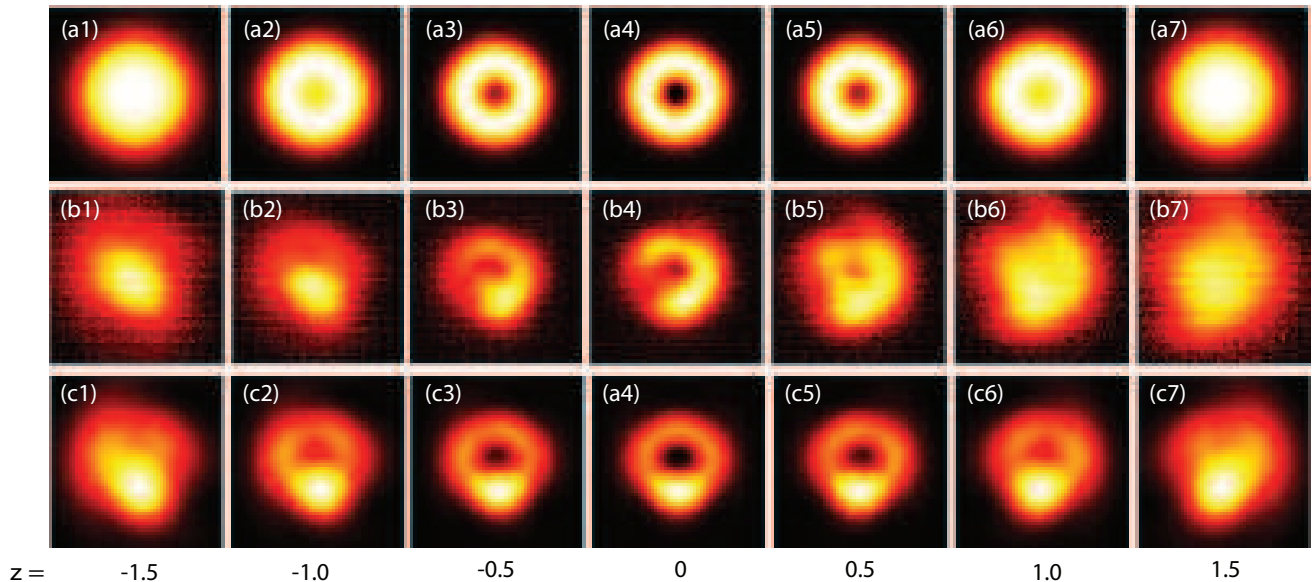}
\caption{(Color online) Experimental (second row) and numerically calculated with Eqs.~(\ref{E_input})--(\ref{E_CRy}) (third row) transverse intensity patterns along the axial direction $z$ for an obstruction with $\phi=45$ for $\rho_0 = 0.92$. The first row shows the numerically calculated transverse intensity patterns obtained in the absence of the obstruction. $z$ is measured in units of the Rayleigh range, which for the Gaussian beam used in our experiments ($w_0 = 44\,\rm{\mu m}$) is $z_R = 9.5\,\rm{mm}$.}
\label{fig7}
\end{figure*}

The evolution of the transverse intensity pattern along the axial direction for a blocking sector of $\phi=45^{\circ}$ is shown in Fig.~\ref{fig7}. Top row shows the transverse intensity patterns in the absence of the blocking mask, while middle and bottom row are the experimental and numerically calculated transverse intensity patterns for the obstructed Gaussian beam. 
Near the focal plane ($z=0$) the transverse intensity pattern differs from the one obtained without obstruction. Additionally, out of the focal plane there is no reconstruction of the transverse intensity pattern. Therefore, for $\rho_0 = 0.92$, the beam reconstruction process is not found at the focal plane nor away from it. 

\section{Conclusions}
\label{conclusions}

We have analyzed in detail the transformation of Gaussian beams partially obstructed when they propagate through a biaxial crystal and parallel to one of the optic axes, i.e., under conditions of conical refraction (CR). We have shown that, at the focal plane, the CR beams for $\rho_0 \gg 1$ preserve the annular shape even when half of the beam is blocked. However, we have found that the dark annular singularity known as Poggendorff dark ring only remains for small perturbations of the input beam. Out of the focal plane we have obtained that the obstruction affects the beam evolution, being its effect more appreciable the further one moves along the axial direction. Additionally, the reconstruction of the state of polarization of the CR beam has also been investigated. For $\rho_0 \gg 1$, we have found that the polarization distribution of the CR rings is very stable against large perturbations.

We have carried out analogous investigations for $\rho_0 \approx 1$. In this case the transverse light pattern is more affected by the presence of the obstruction than in the case of $\rho_0 \gg 1$. Regarding the reconstruction of the state of polarization we have found that only the azimuth of the polarization is relatively robust when large obstructions affect the input Gaussian beam.

\end{document}